%% file: lsdeformed_v3.tex
\documentclass[11pt, oneside]{article}   	

\usepackage{tikz}

\usepackage{jheppub}
\usepackage{comment}

\usepackage[UKenglish]{babel}
\usepackage[T1]{fontenc}
\usepackage{lmodern}

\usepackage{graphicx}
\usepackage{caption}
\usepackage{subcaption}	

\usepackage{amsmath}	
\usepackage{physics}
\usepackage{float}
\usepackage[utf8]{inputenc}
\usepackage{tikz}
\usetikzlibrary{decorations.pathmorphing}
\usetikzlibrary{positioning}
\usetikzlibrary{arrows}
\usetikzlibrary{decorations.markings,arrows.meta,bending}
\usepackage{standalone}

\DeclareMathOperator\Log{Log}

\usepackage{color}
\usepackage{mathtools}
\usepackage{amssymb}
\usepackage{todonotes}

\usepackage{hyperref}

\hypersetup{
    bookmarks=true,
    colorlinks,
    citecolor=blue,
    filecolor=black,
    linkcolor=blue,
    urlcolor=blue
}

\usepackage{overpic}
\usepackage{array}
\usepackage{rotating}

\usepackage{amssymb}

\title{The spectrum of a quantum Lifshitz black hole in two dimensions}

\author{Matthias Harksen$^{1}$ and}
\author{Watse Sybesma$^{2,3}$}
\affiliation{$^{1}$Science Institute, University of Iceland\\ Dunhaga 3, 107 Reykjav\'{i}k, Iceland}
\affiliation{$^{2}$Department of Applied Mathematics and Theoretical Physics\\ University of Cambridge, Cambridge CB3 0WA, United Kingdom}
\affiliation{$^{3}$Isaac Newton Institute for Mathematical Sciences\\ University of Cambridge, Cambridge CB3 0EH, United Kingdom}
\emailAdd{mbh6@hi.is}
\emailAdd{watse.sybesma@su.se}

\abstract{We examine the low-energy spectrum of a four-dimensional near-extremal black hole that arises as a solution to a low energy effective theory of heterotic string theory. 
The effective two-dimensional gravitational description exhibits features of Lifshitz symmetry, which break the usual $\text{SL}(2,\mathbb{R})$ invariance down to $U(1)$. 
For this effective two-dimensional gravitational description, we derive a one-dimensional Schwarzian-like action that inherits the $U(1)$ symmetry.
The Schwarzian-like description allows us to compute a logarithmic correction to the entropy through a saddle-point approximation of the two-dimensional partition function. This logarithmic correction modifies the density of states, lifting the delta-function divergence at extremality, and removes the exponential ground-state degeneracy seen in the semiclassical analysis. Furthermore, the prefactor of the logarithmic term is $\frac{1}{2}$, rather than $\frac{3}{2}$ found for the $\text{SL}(2,\mathbb{R})$ invariant description, indeed reflecting having fewer symmetries.}

\begin{document}
\maketitle
\newpage

\section{Introduction}
In the extremal limit of a Reissner-Nordström black hole the temperature vanishes while the Bekenstein-Hawking entropy remains finite and exponentially large suggesting that an extremal black hole has a large number of ground states compared to the degrees of freedom. Phenomenologically this large entropy at zero temperature is already in tension with Nernst's third law of thermodynamics, which suggests that both the entropy at zero temperature as well as the ground state degeneracy are expected to be small. Page suggested in \cite{Page:2000dk} that a way to resolve this tension was by showing that for extremal black holes all states are non-degenerate while they are separated by exponentially tiny energy gaps.

\begin{figure}[H]
\begin{center}
	\input{puzzle-diagram.tex}
	\vspace{-0.5cm}
\end{center}
\caption{The thermodynamically accessible energy of the black hole above extremality scales with temperature as $E_{\text{BH}} \sim T^{2}$, while according to Wien's displacement law, a typical quanta of Hawking radiation scales linearly with temperature, $E_{\text{Hawking}} \sim T$. The semi-classical description breaks down for $T \leq T_{\text{gap}}$ where $(T_{\text{gap}}, E_{\text{gap}})$ marks the intersection of the two curves. The quantum logarithmic correction (purple line) modifies the energy above extremality to $E_{\text{BH}} \sim \frac{3}{2}T + T^2$, resolving the thermodynamics at small temperatures  \cite{Iliesiu:2020qvm}.}
    \label{fig:puzzle-diagram}
\end{figure}
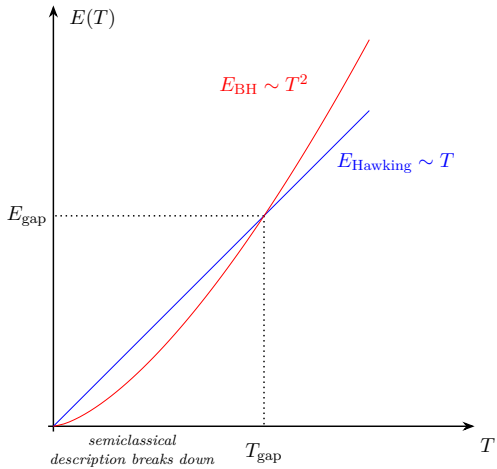

There is a second puzzle referred to as the mass-gap puzzle which was pointed out in \cite{Preskill:1991tb} and further elaborated in \cite{Maldacena:1998uz}. Namely, as the black hole approaches extremality the wavelength of an emitted quanta of Hawking radiation becomes large compared to the size of the black hole. In fact for low values of the temperature the typical emitted quanta of Hawking radiation, $E_{\text{Hawking}}$, exceeds the thermodynamically accessible energy of the black hole above extremality, $E_{\text{BH}}$,
\begin{align} \label{eq:scalingpuzzle}
	E_{\text{BH}} \sim T^{2}, \hspace{1cm} E_{\text{Hawking}} \sim T \,.
\end{align}
The intersection of the two curves $E_{\text{BH}}$ and $E_{\text{Hawking}}$, implies the presence of a conjectured gap above extremality, $E_{\text{gap}}$, below which the semi-classical thermodynamics no longer applies (see \autoref{fig:puzzle-diagram}) since even emitting a single quanta of Hawking radiation would lead to inconsistencies in the semi-classical analysis of the black hole thermodynamics.

Several different aspects of these two\footnote{It has been suggested from a random-matrix model prescription that a large gap will appear whenever there is a mechanism producing a high degree of ground state degeneracy \cite{Johnson:2024tgg, Hernandez-Cuenca:2024icn}.} puzzles have been studied throughout the years and many resolutions have been proposed for determining the energy levels of a charged black hole. Early work on microstate counting for supersymmetric extremal black holes by Strominger and Vafa \cite{Strominger:1996sh} showed that it was possible to reproduce the correct Bekenstein-Hawking entropy from a microscopic prescription and seemed to imply that there indeed was a large ground state degeneracy present at extremality. Furthermore Maldacena and Susskind \cite{Maldacena:1996ds} provided D-brane constructions with gaps between extremal and near-extremal states. Maldacena and Strominger \cite{Maldacena:1997ih} later provided a gravitational derivation of the proposed mass-gap scale purely from supersymmetric Einstein-Maxwell gravity without retorting to string theory arguments. The existence of mass-gap in the case of supersymmetric black holes which preserve two supercharges or more is strongly supported by a vast gallery of examples, see e.g.~\cite{Heydeman:2020hhw, Boruch:2022tno, Iliesiu:2022kny, Iliesiu:2022onk, Sen:2023dps}. Notably, in the case of $\mathcal{N} = 1$ supersymmetric black holes do not exhibit a gapped spectrum \cite{Stanford:2024mek}. Deformations of $\mathcal{N} = 2$ SYK are also able to close the spectral gap \cite{Heydeman:2024ohc}.

Recently it has been argued that the existence of a mass-gap is an artefact of supersymmetry rather than a generic feature of non-supersymmetric black holes \cite{Iliesiu:2020qvm, Turiaci:2023wrh}. Logarithmic corrections to the entropy of extremal black holes imply that by including quantum corrections for the gravitational path integral the spectrum and dynamics of near-extremal black holes is modified such that at the expected gap energy scale there is a continuous spectrum of states \cite{Iliesiu:2020qvm, Heydeman:2020hhw, Iliesiu:2022onk, Banerjee:2023quv, Kapec:2023ruw, Rakic:2023vhv, Maulik:2024dwq}. The techniques for carrying out these computations were developed by Sen and collaborators in \cite{Banerjee:2010qc, Banerjee:2011jp, Sen:2012kpz, Sen:2012cj} for the application to supersymmetric black holes.

 Alternatively, viewing the mass-gap as an artefact of supersymmetry is supported in large part by the extensive work carried out in highlighting the relationship between near-extremal Reissner-Nordström black holes and the Schwarzian theory, which we elaborate on a bit further. In the near-extremal limit of a Reissner-Nordström black hole the metric factorises into $\text{AdS}_2 \times S^2$ and by dimensionally reducing the Einstein-Maxwell action on a transverse two-sphere, an effective two-dimensional description is obtained which captures some of the dynamics of the four-dimensional black hole \cite{Navarro-Salas:1999zer, Nayak:2018qej}.\footnote{For a nice review of this dimensional reduction we refer the reader to section 6 of \cite{Brown:2018bms}.} In this case, this two-dimensional effective theory turns out to be Jackiw-Teitelboim (JT) gravity \cite{Jackiw:1984je, Teitelboim:1983ux}.\footnote{For a more detailed review of JT gravity we refer the reader to \cite{Sarosi:2017ykf, Mertens:2022irh}.} In the full gravitational path integral for JT gravity the dilaton  can be integrated out \cite{Saad:2019lba, Moitra:2021uiv} and by imposing wiggling boundary conditions the path integral reduces to that of the Schwarzian theory \cite{Almheiri:2014cka, Maldacena:2016upp, Engelsoy:2016xyb}. The path integral of the Schwarzian theory is one-loop exact and hence the result represents the full gravitational path integral for JT gravity \cite{Stanford:2017thb} and can be used to compute $\text{log}(T)$ corrections. Note that this two-dimensional description does not capture all $\text{log}(T)$ corrections that can arise in the four-dimensional partition function, as it ignores modes that are related to the two-sphere and the the gauge field.
 This Schwarzian theory also arises in the low energy effective limit of SYK \cite{Maldacena:2016hyu, Polchinski:2016xgd, Cotler:2016fpe, Saad:2018bqo}, which gives rise to a holographic duality between JT gravity and the low energy dynamics of the SYK model, see e.g. \cite{Saad:2019lba}.

The purpose of this paper is to consider the mass-gap puzzle for non-supersymmetric Lifshitz black holes that arise in a low energy effective description of string theory and compute the logarithmic corrections to the entropy. In Lifshitz theories the metric displays an anisotropic scale invariance between the temporal and spatial coordinates in stark contrast to the relativistic scale invariance \cite{Kachru:2008yh, Horowitz:2011gh}, reducing the number of possible isometries. We aim to explicitly investigate the notion that the $\log(T)$ corrections to the black hole entropy depend on the amount of symmetry emerging near the horizon in the small temperature limit, as has been argued in, e.g., \cite{Heydeman:2020hhw, Iliesiu:2020qvm}. In passing we derive a one-dimensional boundary description of the effective two-dimensional theory of gravity that we will employ, which explicitly breaks conformal symmetry and as such can be the starting point of studying non-conformal incarnations of low-dimensional holography.

The rest of the paper is organised as follows. In section \ref{sec:CMLS} we show the intricate connection between a four-dimensional Lifshitz black hole and a one-dimensional Lifshitz-like boundary action which shares similar features to that of the Schwarzian. This is done in the following manner; in subsection \ref{subsec:4d} we introduce the four-dimensional Lifshitz black hole action and study its near-horizon geometry. We show that the geometry factorises into $\text{Lif}_{2,\alpha} \times S^2$ where $\text{Lif}_{2,\alpha}$ is a Lifshitz-like spacetime that is notably asymptotically $\text{AdS}_2$ for Lifshitz parameter $0 < \alpha < 1$. We find that the thermodynamically accessible energy of the black hole above extremality scales with temperature as $E_{\text{BH}} \sim T^{1+\frac{1}{\alpha}}$. This leads to the same mass-gap puzzle for these Lifshitz black holes as we highlighted previously in \eqref{eq:scalingpuzzle} for near-extremal Reissner-Nordström black holes. In subsection \ref{subsec:2d} we dimensionally reduce the four-dimensional model and study its two-dimensional counterpart. We summarise the main features of the resulting semi-classical thermodynamics in subsection \ref{subsec:semi-thermo}. In subsection \ref{subsec:boundary-action} we derive a Lifshitz-like boundary action which captures the same semi-classical thermodynamics as the two-dimensional Lifshitz-like dilaton model. In section \ref{sec:QBH} we compute the quantum corrected partition function and the corresponding density of states. We compute the $1$-loop \emph{approximation} to the quantum corrected partition function with a saddle point approximation in subsection \ref{subsec:qpartition} and derive the corresponding density of states in subsection \ref{sec:dos} while highlighting the implications of our findings for the mass-gap puzzle for non-supersymmetric black holes which break conformal symmetries.

\section{A Lifshitz-like dilaton gravity model} \label{sec:CMLS}
We start with considering the decoupling limit of the four-dimensional Lifshitz model under consideration in subsection \ref{subsec:4d} and are interested in computing loop corrections.
In order to compute loop corrections in the decoupling limit, we turn to the effective two-dimensional description that arises in the decoupling limit, of which we study the geometry and thermodynamics in subsections \ref{subsec:2d} and \ref{subsec:semi-thermo}, respectively. Finally, we derive a boundary description for the two-dimensional Lifshitz-like model in subsection \ref{subsec:boundary-action}, which will allow us to compute a loop correction.
\subsection{The four-dimensional model} \label{subsec:4d}
In this subsection we consider the following four-dimensional action presented in \cite{Cadoni:1993yt, Cadoni:1993rn} which was derived in from the low energy limit of a dimensional reduction of a ten-dimensional heterotic supergravity theory using a suitable truncation of the spectrum in \cite{Witten:1985xb},
\begin{align}\label{eq:4daction}
	I = \int d^4x \sqrt{-g} \, e^{-2\phi} \left[ R + 4 \left( \nabla \phi \right)^2 - \frac{2}{3} (\nabla \psi)^2 - F^2 - e^{2(\phi - \tfrac{q}{3}\psi)} F^2 \right]\,.
\end{align}
We will refer to $\phi$ as dilaton, $\psi$ as compacton, and $F_{\mu\nu} = \nabla_\mu A_\nu - \nabla_\nu A_\mu$ the Maxwell field strength, with $q$ a coupling constant. The equations of motion for the action presented in \eqref{eq:4daction} are given by
\begin{subequations} \label{eq:4d-eom}
    \begin{align}
        R + 4\nabla^2 \phi - 4 (\nabla \phi)^2 - \frac{2}{3} (\nabla \psi)^2 - F^2 &= 0, \label{eq:4d-eom-dilaton} \\
        \nabla^2 \psi - 2 \nabla^\mu \phi \nabla_\mu \psi + \frac{q}{2} e^{2(\phi - \frac{q}{3} \psi)} F^2 &= 0, \label{eq:4d-eom-compacton} \\
        \nabla_\mu \left[ \left( e^{-2\phi} + e^{-\frac{2q}{3} \psi} \right) F^{\mu\nu} \right] &= 0, \label{eq:4d-eom-maxwell} \\
        R_{\mu\nu} - \frac{1}{2} R g_{\mu\nu} + 2 \nabla_\mu \nabla_\nu \phi - 2 \nabla^2 \phi \, g_{\mu\nu} + 2 (\nabla \phi)^2 g_{\mu\nu} - \frac{2}{3} \nabla_\mu \psi \nabla_\nu \psi & \notag \\
         + \frac{1}{3} (\nabla \psi)^2 g_{\mu\nu} + \frac{1}{2} \Big(1 + e^{2(\phi - \frac{q}{3} \psi)} \Big) F^2 g_{\mu\nu}
        - 2 \Big(1 + e^{2(\phi - \frac{q}{3} \psi)} \Big) F_{\mu\rho} F_\nu{}^\rho &= 0, \label{eq:4d-eom-metric}
    \end{align}
\end{subequations}
where the first corresponds to varying the dilaton, the second with varying the compacton, the third corresponds with varying the Maxwell-field $A_{\mu}$ and the last equation corresponds to varying the metric. An exact magnetically charged black hole solution is given by \cite{Cadoni:1993yt, Cadoni:1993rn}
\begin{equation} \label{eq:4d-solution}
    \begin{aligned}
        ds^2 &= -\left(1-\frac{r_+}{r}\right) \left(1 - \frac{r_-}{r} \right)^{\frac{-1+\alpha}{1+\alpha}} dt^2 + \left(1-\frac{r_+}{r}\right)^{-1} \left(1-\frac{r_-}{r}\right)^{-1} dr^2 + r^2 d\Omega^2\,, \\
        \phi(r) &= -\frac{1}{2(1+\alpha)} \ln \left(1-\frac{r_-}{r}\right), \quad
        \psi(r) = \frac{3}{q} \left[ \phi(r) - \frac{1}{2} \ln \left(2 \alpha \right) \right], \quad
        A = Q_m \cos\theta \, d\phi\,,
    \end{aligned}
\end{equation}
where we defined $\alpha = \frac{3}{2q^2} \geq 0$ and the two parameters $r_+$ and $r_-$ correspond to outer and inner horizons respectively and can be related to the magnetic monopole charge $Q_m$ and the mass $M$ of the black hole by the following relation
\begin{align}
	M = \frac{1}{2}r_+ + \frac{1}{2}\frac{\alpha}{1+\alpha}r_-\,,
	\quad Q_m^2 = \frac{1}{2}\frac{r_{+}r_{-}}{1+\alpha}\,.
\end{align}
By investigating the Hawking temperature
\begin{equation}
	T
	=
	\frac{1}{4\pi r_{+}}
	\left(
		1
		-
		\frac{r_{-}}{r_{+}}
	\right)^{\frac{\alpha}{1+\alpha}}
	\,,
\end{equation}
we conclude that $r_0=r_{+}=r_{-}$ leads to an extremal black hole.

\paragraph{Decoupling regime.}
Let us investigate the near-extremal decoupling limit for $0 < \alpha < 1$. By fixing the charge to its extremal value
\begin{align}
	Q_{0} = \frac{r_0}{\sqrt{2(1+\alpha)}}\,.
\end{align}
The horizons $r_\pm(T)$ have the following small temperature expansion at lowest order
\begin{align} \label{eq:rtexp}
	r_\pm(T) = r_0 \pm \frac{r_0}{2 }\left(4 \pi r_0 T\right)^{1+ \frac{1}{\alpha}} + \ldots \, ,
\end{align}
from which we find that the mass near extermality is given by the following expression to the lowest order in temperature
\begin{equation}
	M(T) = M_0 + \frac{T^{1 + \frac{1}{\alpha}}}{M_{\text{gap}}}
	\,,
\end{equation}
where the extremal mass and the mass gap are respectively given by
\begin{equation}
	M_0 :=\frac{1 + 2\alpha}{1+\alpha} \frac{r_0}{2}\,,
	\quad
	\frac{1}{M_{\text{gap}}}
	:=
	\frac{1}{1+\alpha} \frac{r_0}{4} (4\pi r_0 )^{1 + \frac{1}{\alpha}}
	\,.
\end{equation} 
Let us now study the decoupling limit of the metric. On top of using the small temperature expansion for $r_+$ and $r_-$ from equation \eqref{eq:rtexp}, to ensure Schwarzschild gauge at leading order we introduce new coordinates $\tilde{t}$ and $\tilde{r}$ such that
\begin{equation}
	t
	=
	\frac{
		\tilde{t}
	}{
		2\pi T
	}
	\,,
	\quad
r = r_+ + r_0  \left[ \left( \frac{\tilde{r}}{\tilde{r}_h} \right)^{1 + \frac{1}{\alpha}} -1 \right]
\left( 4\pi r_0 T \right)^{1+\frac{1}{\alpha}}
	\,,
\end{equation}
where the horizon location in the new radial coordinate is indicated by the value
\begin{equation}
	\tilde{r}_h = 2 r_0^2 \left(1 + \frac{1}{\alpha} \right)
	\, .
\end{equation}
In the small-temperature regime, $T \ll 1$, the new radial coordinate satisfies $\frac{\tilde{r}}{\tilde{r}_h}  \sim \frac{1}{4\pi r_0 T} \gg 1$, consistent with the near-horizon approximation. In these coordinates, the metric reads\footnote{Although not required for the analysis presented here, the next order correction is given by
\begin{align}
	\frac{ds^2}{(4 \pi r_0 T)^{1+\frac{1}{\alpha}}} =& - \frac{2 \alpha^2 \tilde{r}_h}{(1+\alpha)^2} \left[ 3\left( \frac{\tilde{r}}{\tilde{r}_h} \right)^2 - \left( \frac{\tilde{r}}{\tilde{r}_h} \right)^{1-\frac{1}{\alpha}} - 2 \left( \frac{\tilde{r}}{\tilde{r}_h} \right)^{3 + \frac{1}{\alpha}} \right]d\tilde{t}^2 \nonumber
	\\&+ \frac{1}{2r_0^2}\left[ \frac{1}{2} - \left( \frac{\tilde{r}}{\tilde{r}_h} \right)^{1+\frac{1}{\alpha}}\right]\left[\left( \frac{\tilde{r}}{\tilde{r}_h} \right)^{2} - \left( \frac{\tilde{r}}{\tilde{r}_h} \right)^{1-\frac{1}{\alpha}} \right]^{-1} d\tilde{r}^2 + r_0^2  \left[-1 + 2\left( \frac{\tilde{r}}{\tilde{r}_h} \right)^{1+\frac{1}{\alpha}} \right] d\Omega^2 \, .\nonumber
\end{align}
}
\begin{equation}
	ds^{2}
	=
	-F(\tilde{r}) d\tilde{t}^{2}
	+
	\frac{d\tilde{r}^{2}}{F(\tilde{r})}
	+
	r_{0}^{2}
	d\Omega_{2}^{2}
	+
	\mathcal{O}(T^{1+\frac{1}{\alpha}}) \,,
\end{equation}
where
\begin{equation}\label{eq:bhfixed}
	F(\tilde{r})
	=
	\frac{2\alpha \tilde{r}_{h}}{1+\alpha} 
	\left[
		\left(
			\frac{\tilde{r}}{\tilde{r}_{h}}
		\right)^{2}
		-
		\left(
			\frac{\tilde{r}}{\tilde{r}_{h}}
		\right)^{1-\frac{1}{\alpha}}
	\right] 
	\,.
\end{equation}
Notice that we cannot turn off the black hole and its temperature is fixed at $\tilde{T}=1/(2\pi)$, which is to be expected for the rescaled time coordinate $\tilde{t}$.
When focussing on the $\tilde{r}-\tilde{t}$ plane the Ricci scalar $R_{\tilde{r}-\tilde{t}}$ reads
\begin{equation}
	R_{\tilde{r}-\tilde{t}}
	=
	\frac{2\alpha}{1+\alpha}
	\frac{1}{\tilde{r}_{h}}
	\left[	
		-
		2
		+
		\frac{1-\alpha}{\alpha^{2}}
		\left(
			\frac{\tilde{r}_{h}}{\tilde{r}}
		\right)^{1+\frac{1}{\alpha}}
	\right]
	\,,
\end{equation} 
which implies that $\alpha=1$ yields a constant negatively valued Ricci scalar and the effective two-dimensional theory is JT gravity. For $0<\alpha<1$ the Ricci scalar is running, but asymptotically always a patch of $\text{AdS}_{2}$. At extremality, we find the following factorisation of space
\begin{equation}
	ds^{2}
	=
	\text{Lif}_{2,\alpha}
	\times
	S^{2}
	\,,
\end{equation}
where $S^{2}$ refers to a two-sphere and $\text{Lif}_{2,\alpha}$ represents a two-dimensional Lifshitz-like spacetime with dynamical critical exponent $\alpha$, which we will discuss in more details in the next section. For now it is relevant to highlight the fact that away from $\alpha=1$ the Ricci scalar runs.

\paragraph{The low temperature puzzle.} The available thermodynamic energy above extremality is now given by $E_{\text{BH}}(T) = M(T) - M_{0} = \frac{1}{M_{\text{gap}}} T^{1+\frac{1}{\alpha}}$. We therefore run into the same mass-gap puzzle as highlighted in the introduction for Reissner-Nordström black holes \cite{Preskill:1991tb}. For low enough values of the temperature the typical emitted quanta of Hawking radiation, $E_{\text{Hawking}} \sim T$ starts to exceed the available energy of the black hole above extremality, $E_{\text{BH}}$,
\begin{equation}
	E_{\text{4d black hole}}\sim T^{1+\frac{1}{\alpha}}
	\,,
	\quad
	E_{\text{Hawking}}
	\sim T
	\,.
\end{equation} 
When these energies become of comparable size, the amount of energy released through Hawking radiation becomes significant compared to the energy available to the black hole. As such the assumption of equilibrium is invalid and for low enough temperatures the saddle should receive large semi-classical corrections.

It was in \cite{Sen:2012cj} where such corrections were computed through considering a set of normalisable zero modes at extremality, composed of the asymptotic Killing vectors. The resulting outcome is an extra term $\sim \log T$ that enters the free energy and modifies the low energy behaviour, averting the above scenario. To apply the method of \cite{Sen:2012cj}, we require the computation of asymptotic Killing vectors for the $\text{Lif}_{2,\alpha}$ spacetime we find at extremality. It turns out that we are unable to do this, so instead we will compute the logarithmic correction through the boundary description of the two-dimensional $\text{Lif}_{2,\alpha}$ theory.
\subsection{The two-dimensional model} \label{subsec:2d}
The effective two-dimensional description of the four-dimensional model presented in \eqref{eq:4daction} is obtained by performing a reduction over a constant two-sphere and integrating out the Maxwell field \cite{Cadoni:1993yt}. Specifically, we insert the ansatz, $\psi(r) = \frac{3}{q} \phi(r) - \frac{3}{2q}\Log(2\alpha)$, for the compacton into the action and reduce over a sphere of constant radius $r_0 = r_+ = r_-$, which corresponds to the extremal radius of the black hole. By factorizing the spacetime metric as
\begin{align}
    g_{\mu \nu}dx^\mu dx^\nu = \tilde{g}_{ab} d\tilde{x}^{a} d\tilde{x}^b + r_0^2 d\Omega_2^2\,,
\end{align}
we obtain the intermediate two-dimensional action
\begin{align}
	I_{\text{2d}} = 4 \pi r_0^2 \int d^2 x \sqrt{-\tilde{g}} \, e^{-2\phi} \left[ \tilde{R} + 4(1-\alpha) (\tilde{\nabla} \phi)^2 + \frac{2}{r_0^2} - 2(1+2\alpha) \frac{Q_0^2}{r_0^4}  \right]\,,
\end{align}
where $\phi$ is the dilaton, $\tilde{R}$ is the two-dimensional Ricci-scalar and $Q_0 = r_0/\sqrt{2(1+\alpha)}$ represents the extremal value of the magnetic monopole charge. By defining the energy scale
\begin{align}
	\lambda = \frac{1}{2 r_0 \sqrt{1+\alpha}}\,,
\end{align}
we can express the action in its final form (dropping the cumbersome tilde notation)
\begin{align}
    I_{\text{CM}} = 4\pi r_0^2 \int d^2x \sqrt{-g} \, e^{-2\phi} \left[ R + 4(1-\alpha) (\nabla \phi)^2 + 4\lambda^2 \right]\,.
\end{align}
While this $4\pi r_0^2$ prefactor naturally arises from the dimensional reduction process and reflects the relationship between the 4D and 2D Newton's constants ($G_4$ and $G_2$), we will instead use the form of the action introduced by Lemos and Sá \cite{Lemos:1993py}, written as\footnote{By introducing $\Phi = e^{-2\phi}$ and Weyl rescaling the metric $g_{ab} \to \Phi^{\alpha -1} g_{ab}$ we can eliminate the kinetic term
\begin{align} \label{eq:flat-CMLS-action}
    I_{\text{W-Lif}} = \frac{1}{2}\int_{\mathcal{M}} d^2 x \sqrt{-g}  \left( R \Phi + 4\lambda^2 \Phi^{\alpha} \right)\,,
\end{align}
This action has been studied in \cite{Mignemi:1994wg}, but does not allow for the asymptotically $\text{AdS}_{2}$ solutions we aim to study.}
\begin{align} \label{eq:cmls-bulk-action}
    I_{\text{Lif}} = 
    \frac{1}{2}\int_{\mathcal{M}} d^2x \sqrt{-g} \, e^{-2\phi} \left[ R + 4(1-\alpha) (\nabla \phi)^2 + 4\lambda^2 \right]
    \,.
\end{align}
We will refer to this as the Lifshitz-like dilaton model, motivated by the relation between entropy and temperature, $S \sim T^{1/\alpha}$, which we will establish in \eqref{eq:entropy}. While in this paper we derive this model from the dimensional reduction of \eqref{eq:4daction}, an alternative derivation was presented in \cite{Sybesma:2022nby}. There, the model arises by considering the near-horizon region of a Lifshitz black hole constructed using bottom-up methods by Taylor \cite{Taylor:2008tg, Taylor:2015glc}, with a large number of transverse dimensions. This connection with Lifshitz black hole models provides another reason for referring to it as Lifshitz-like. Let us now describe the significance of the dimensionless parameter $\alpha$, which determines the asymptotics of the solutions:
\begin{itemize}
    \item \underline{$\alpha = 1$:} The action reduces to the Jackiw-Teitelboim (JT) model \cite{Jackiw:1984je, Teitelboim:1983ux}, which describes asymptotically $\text{AdS}_2$ spacetimes and plays a central role in near-horizon black hole physics and holography.
    \item \underline{$\alpha = 0$:} The action corresponds to the Callan-Giddings-Harvey-Strominger (CGHS) model \cite{Callan:1992rs}, which admits asymptotically flat solutions. These solutions are not the focus of our study and will not be considered further.
    \item \underline{$0 < \alpha < 1$:} The model admits asymptotically $\text{AdS}_2$ solutions but with modified scaling properties, leading to a Lifshitz-like behavior. This range of $\alpha$ interpolates between the JT and CGHS models and is the primary focus of our analysis.
\end{itemize}
Varying the action in \eqref{eq:cmls-bulk-action} with respect to the metric and dilaton yields respectively
\begin{subequations} \label{eq:LS-EE}
    \begin{eqnarray}
        \nabla_\mu \nabla_\nu \phi - 2 \alpha \nabla_\mu \phi \nabla_\nu \phi  - \left( \nabla^2 \phi - (1+\alpha) (\nabla \phi)^2 + \lambda^2 \right) g_{\mu \nu}  & = 0\,, \label{eq:LS-EE-metric} \\
    R + 4(1-\alpha) \nabla^2 \phi - 4(1-\alpha)(\nabla \phi)^2 + 4\lambda^2 &= 0\,. \label{eq:LS-EE-dilaton}
\end{eqnarray}
\end{subequations}
The general solution for $0 < \alpha < 1$ in Schwarzschild gauge is given by
\begin{align} \label{eq:LS-sol}
	ds^{2}
	=
	-f(r)dt^{2}
	+
	\frac{dr^{2}}{f(r)}
	\,,
	\quad
   	f(r) = (ar)^2 - b(ar)^{1-\frac{1}{\alpha}}\,, 
	\quad
	\phi(r) = -\frac{1}{2\alpha}\ln(ar)\,,
\end{align}
where the constant $a$ is just a convenient redefinition of $\lambda$ given by
\begin{align}
    a = \frac{2\alpha \lambda}{\sqrt{1+\alpha}}\geq 0\,,
\end{align}
and $b$ is a free parameter which, from a higher-dimensional perspective, parametrizes the departure of the solution from extremality \cite{Cadoni:1993rn}. In the context of the dimensional reduction, the parameters $a$ and $b$ can be read off from \eqref{eq:bhfixed} in terms of the extremal radius $r_0$ of the higher-dimensional black hole, where $a = \frac{\alpha}{r_0(1+\alpha)}$ and $b = (2 r_0)^{1 + \frac{1}{\alpha}}$. For this section, however, we will treat $b$ as an unconstrained free parameter, reflecting the freedom inherent in the two-dimensional dilaton model. Defining the horizon radius $r_h$ implicitly for the two-dimensional model by $f(r_h) = 0$, we have additionally the following relation $b = (a r_h)^{1+\frac{1}{\alpha}}$. 

We can relate the free parameter $b$ to the ADM mass $M$ of the solution \cite{Cadoni:1993rn, Lemos:1993py, Kumar:1994ve}.
\begin{align}\label{eq:ADM_mass}
	M = \frac{ab}{2 \alpha}
	\,.
\end{align}
Additionally it may be verified by using the equations of motion \eqref{eq:LS-EE-metric} that
\begin{equation}
	\xi^{\mu} = \epsilon^{\mu \nu} \nabla_\nu (e^{-2\alpha \phi})
	\,,
\end{equation}
is always a Killing vector of the metric which can be used to compute the aforementioned ADM mass \cite{Mann:1992yv}. In Schwarzschild gauge this becomes $\xi
	=
	-a\partial_{t}$. The existence of this Killing vector for a dynamical dilaton already indicates how the $\text{SL}(2,\mathbb{R})$ isometries are broken down to $U(1)$ and hence from $\text{AdS}_2$ to nearly-$\text{AdS}_2$ in the case of JT gravity \cite{Maldacena:2016upp}.

The Ricci scalar of the metric for $0 < \alpha < 1$ is given by
\begin{align}
	R = -2a^2 - 2a M \left( 1 - \tfrac{1}{\alpha} \right) (ar)^{-(1+\frac{1}{\alpha})}\,.
\end{align}
From this Ricci scalar we see that the solution for large radial values is asymptotically $\text{AdS}_2$. We may additionally defer from this that for $b = 0$, i.e.,~$M= 0$ the solution is pure $\text{AdS}_2$.

An important difference to highlight between $\alpha=1$ (JT gravity) and $0 < \alpha < 1$ is how a dynamical dilaton affects the isometries. For JT one can toggle $\exp \phi=0$, independent of the metric, such that the bulk reflects pure $\text{AdS}_{2}$ solutions. Turning on the dilaton breaks down the $\text{SL}(2,\mathbb{R})$ symmetry to $U(1)$, so-called nearly $\text{AdS}_{2}$ \cite{Maldacena:2016upp}.

To conclude this section we highlight the Lifshitz-like nature of this dilaton model. The line element presented in \eqref{eq:LS-sol} is invariant under the following Lifshitz scale transformation
\begin{align} \label{eq:lifshitzscale}
	t\to L^{-1} t
	\,,
	\quad
	r\to L r
	\,,
	\quad
	b\to L^{1+\frac{1}{\alpha}}b
	\,.
\end{align}
This anisotropic scale invariance between the temporal and spatial coordinates is in stark contrast to the conformal scale invariance and is a general hallmark of Lifshitz theories \cite{Kachru:2008yh, Horowitz:2011gh}. Furthermore, under the Lifshitz scale transformation the action \eqref{eq:cmls-bulk-action} transforms as
\begin{align}
	I_{\text{Lif}} \to L^{\frac{1}{\alpha}}I_{\text{Lif}}\,,
\end{align}
mainly as an artefact of the dilaton scaling as $e^{-2\phi} \to L^{\frac{1}{\alpha}} e^{-2\phi}$. For $\alpha \neq 1$ the action is therefore not invariant but rather a \emph{similarity} of the action \cite{Biggs:2023sqw}. This is also sometimes referred to as a \emph{hyperscaling violation} of the action \cite{Fisher:1986zz,Huijse:2011ef}. The hyperscaling violation exponent $\frac{1}{\alpha}$ therefore naturally shows up in the thermodynamical analysis of the model. In fact the Lifshitz scale tranformation in equation \eqref{eq:lifshitzscale} implies that the inverse temperature changes as $\beta \to L^{-1} \beta$. Therefore the temperature changes as $T \to LT$ according to \eqref{eq:lifshitzscale} and this in turn means that the entropy scales as $S \to L^{\frac{1}{\alpha}}S$. Since the entropy is a function of the temperature we have shown by scaling arguments that entropy will be proportional to $S \sim T^{\frac{1}{\alpha}}$ which we will formally derive in the next section. 

\subsection{Thermodynamics of the two-dimensional bulk theory} \label{subsec:semi-thermo}

The thermodynamical properties of the solution given by \eqref{eq:LS-EE-metric} have previously been analysed in \cite{Cadoni:1993rn, Kumar:1994ve, Sybesma:2022nby}. In this section we summarise succinctly these thermodynamical properties. Let us first supplement the bulk action \eqref{eq:cmls-bulk-action} with the corresponding Gibbons-Hawking-York boundary term which is needed to make the gravitational variational problem well-defined. Additionally we derive a counterterm that renders the Euclidean on-shell action finite. Including both these boundary terms the action may be written as
\begin{equation}
\begin{aligned} \label{eq:fancy-lif-action}
    I_{\text{Lif}} =&~
    -
    \frac{1}{2}\int_{\mathcal{M}} d^2x \sqrt{g} \, e^{-2\phi} \left[ R + 4(1-\alpha) (\nabla \phi)^2 + 4\lambda^2 \right]
    \\
    &
    -
    \int_{\partial \mathcal{M}} du \sqrt{h}\, e^{-2\phi}\left[ K - \frac{2 \lambda }{\sqrt{1+\alpha}}\right],
\end{aligned}
\end{equation}
where $K$ denotes the trace of the extrinsic curvature and and $h = h_{uu}$ is the induced metric on the boundary parameterised by the coordinate $u$. The free energy, $F$, is then given by
\begin{align} \label{eq:free-energy}
	F = \frac{\mathcal{I}_{\text{Lif}}}{\beta} 
	= -\alpha M
	 = -\frac{2 \pi \alpha }{1 + \alpha} \left( \frac{ \pi a}{\alpha \lambda^2} \right)^{\frac{1}{\alpha}} T^{1+\frac{1}{\alpha}}
	 \,,
\end{align}
where $\beta = \frac{1}{T}$ denotes the inverse of the Hawking temperature of the black hole and $M$ denotes the ADM mass \eqref{eq:ADM_mass}.  
The entropy can then be computed from the relationship
\begin{align} \label{eq:entropy}
	S(T) = -\frac{\partial F}{\partial T} = 2\pi \left( \frac{\pi a}{\alpha \lambda^2} \right)^{\frac{1}{\alpha}} T^{\frac{1}{\alpha}}\,.
\end{align}
This equation highlights the Lifshitz-like nature of the model due to the scaling relation between temperature and entropy. \newpage

For future reference we rewrite the entropy in terms of the ADM mass
\begin{align}
	S(M) = 2\pi \left( \frac{a}{2\alpha} \right)^{- \frac{1}{1+\alpha}} M^{\frac{1}{1+\alpha}}\,.
\end{align}
Through the first law, $F=E-ST$, we can now verify that indeed energy equals to the ADM mass, $E=M$. 
We are in particular able to verify the following identity
\begin{align}
	TS = (1+\alpha )E
	\,.
\end{align}
which is a two-dimensional incarnation of the Smarr formula \cite{Smarr:1972kt}.

\subsection{Deriving the boundary description of the two-dimensional model} \label{subsec:boundary-action}
In this section we will compute the boundary description of the Lifshitz-like model that reflects the $U(1)$ symmetry this model has. 
We note that in $\alpha=1$ case, the usual Schwarzian boundary theory is modified by turning on an explicit source, breaking down the symmetries to $U(1)$.

Let us review the idea behind the Schwarzian computation in JT gravity so that we have a better grip on performing the equivalent computation for the Lifshitz-like model. The Schwarzian description arises from considering a Euclidean disc of the $\text{AdS}_{2}$ vacuum and studying the reparametrisation freedom of the disc.
The $\text{SL}(2,\mathbb{R})$ freedom of the Schwarzian reflects the existence of a vacuum in which the metric equals $\text{AdS}_{2}$, which is $\text{SL}(2,\mathbb{R})$ invariant, and where the dilaton is constant. If the constant dilaton were running, say dependent on the radial coordinates, then $\text{SL}(2,\mathbb{R})$ gets spontaneously broken to $U(1)$.
In the case of JT, all metric solutions are always $\text{AdS}_{2}$ due to the Ricci scalar always equalling the same constant on-shell. 
For that reason one can perform a coordinates transformation to go between the $\text{AdS}_{2}$ vacuum patch and the `black hole' patch once the Schwarzian is obtained.
In the Lifshitz-like model we have access to two distinct solutions, $\text{AdS}_{2}$, and a black hole patch where the Ricci scalar is running. The $\text{AdS}_{2}$ vacuum patch and black hole patch are only related through a large diffeomorphism.
Contrasting the JT case, even the $\text{AdS}_{2}$ vacuum has to be supported by a running dilaton. As such, $U(1)$ symmetry is the maximally attainable symmetry in both cases. When performing the boundary computation we will start from the black hole patch. 
Let us first rewrite the bulk as
\begin{align}\label{eg:bulkcontribution}
    \begin{split}
   I_{\text{bulk}} &= -\frac{1}{2}\int_{\mathcal{M}} d^2x \sqrt{g} \, e^{-2\phi} \left( R + 4(1-\alpha) (\nabla \phi)^2 + 4\lambda^2 \right)  \\ &= -\frac{1}{2} \int_\mathcal{M} d^2x \sqrt{g} \, e^{-2\phi} \left(  8(1-\alpha) (\nabla \phi)^2 - 4(1-\alpha) \nabla^2\phi \right) \\
   &= 2(1-\alpha)\int_{\mathcal{M}} d^2x \sqrt{g} \, \nabla_\mu \left( g^{\mu \nu} e^{-2\phi} \nabla_\nu \phi \right) \\
    &= 2(1-\alpha) \int_{\partial \mathcal{M}} dt \sqrt{h} \,  e^{-2\phi}n^{\mu} \nabla_\mu \phi 
\,,
    \end{split}
\end{align}
where to go to the second line we used the dilaton equation of motion and to go to the third line we used $ e^{-2\phi} \nabla_\mu \nabla_\nu \phi 
    = \nabla_\mu \left( e^{-2\phi} \nabla_\nu \phi \right) + 2e^{-2\phi} \nabla_\mu \phi \nabla_\nu \phi$. In the last line we introduced the unit outward normal $n^\mu$ on $\partial \mathcal{M}$. We flesh these manipulations out in great detail in order to highlight that at no stage did we need to neglect any additional boundary terms. Finally, let us additionally remark that the contribution from evaluating this bulk piece on the horizon vanishes. With this in mind we may rewrite the entire action in the following boundary form
\begin{align}
    I_{\text{Lif}} = &
    -\int_{\partial \mathcal{M}} dt \sqrt{h}  \, e^{-2\phi} \left( -2(1-\alpha)n^\mu \nabla_\mu \phi + K - \frac{a}{\alpha} \right)
    \,.
\end{align}
We will now wick rotate the $t$ coordinate and consider an enclosed boundary parametrised by some periodic coordinate $u$
\begin{equation}\label{eq:ranges}
	u\sim u+u_{\text{max}}
	\,,
	\quad
	t(u+u_\text{max})
	=
	t(u)
	+
	\gamma
	\,.
\end{equation}	
Note that when $\gamma\neq\beta$, where $\beta$ is the inverse Hawking temperature, we have a conical singularity. 
The metric induced on the boundary $h_{uu}$ is given by
\begin{equation}\label{eq:boundary_size}
	h_{uu}(t(u),r(u))
	=
	f(r(u))(t'(u))^{2}
	+
	\frac{(r'(u))^{2}}{f(r(u))}
	=
	\frac{1}{\epsilon^{2}}
	\,,
\end{equation}
where $|\epsilon|\ll 1$ and where we chose $1/\epsilon^{2}$ as we expect the leading order to coincide with $\text{AdS}_{2}$ behaviour. 
With $f$ we denote the emblackening factor presented in \eqref{eq:LS-sol}.
It will turn out towards the end of the computation that we need to rescale $t(u)$, $\gamma$ and the black hole mass parameter $b$ in order to retain a non-trivial solution for $\alpha\neq1$. 
From \eqref{eq:boundary_size} we solve
\begin{equation}
	r(u)
	=
	\frac{1}{a\epsilon t'(u)}
	+
	\mathcal{O}(\epsilon^{3})
	\,,
\end{equation}
where we point out that in $f(r)$ for large radial values $(ar)^2$ will dominate $b(ar)^{1-\frac{1}{\alpha}}$ since $0 < \alpha < 1$.
Let us now turn to computing the extrinsic curvature as a function of $u$. 
For a curve $(t(u),r(u))$ the normal is given by
\begin{equation}
	n_{\mu} = \frac{1}{\sqrt{h_{uu}}}(-r'(u),t'(u))
	\,.
\end{equation}
Introducing the vielbein $e^{\mu}_{u}=(t'(u),r'(u))$ we can evaluate the extrinsic curvature as
\begin{equation}
	K
	=
	h^{uu}K_{uu}
	=
	h^{uu}e^{\mu}_{u}e^{\nu}_{u}
	(\nabla_{\mu}n_{\nu})
	\,,
	\quad
	\Gamma^{t}_{tr}=-\Gamma^{r}_{rr}=-\frac{1}{2}\partial_{r}f(r)
	\,,
	\quad 
	\Gamma^{r}_{tt}=\frac{1}{2}f(r)\partial_{r}f(r)
	\,.
\end{equation}
We can now present
\begin{align}
    K &= 
    a 
    + 
    \frac{1}{a} \{t,u\} \epsilon^2 
    + 
    E (t'(u) \epsilon)^{1+\frac{1}{\alpha}}
	+ 
	\mathcal{O}(\epsilon^{4})
    \,,  
    \\
    \sqrt{h} &=  
    \frac{1}{\epsilon} 
    + \frac{\epsilon}{2a^2} \left(\tfrac{t''(u)}{t'(u)}\right)^2 - \frac{\alpha E}{a} t'(u)^{1+\frac{1}{\alpha}} \epsilon^{\frac{1}{\alpha}} +
    \mathcal{O}(\epsilon^{3})\,, 
    \\
    e^{-2\phi} &= (ar(u))^{\frac{1}{\alpha}} =
   \left( \tfrac{1}{\epsilon t'(u)} \right)^{\frac{1}{\alpha}} + \mathcal{O}(\epsilon^{4-\frac{1}{\alpha}})\,, \label{eq:boundarycond}
    \\
    -2(1-\alpha)n^\mu \nabla_\mu \phi &=
    \left(\tfrac{1}{\alpha} - 1 \right) a 
    - 
    \tfrac{1-\alpha}{2 a \alpha} \left( \tfrac{t''(u)}{t'(u)} \right)^2\epsilon^{2}  -E(1-\alpha ) (t'(u) \epsilon)^{1+\frac{1}{\alpha}}
    + \mathcal{O}(\epsilon^4)\,,
\end{align}
where we introduced the Schwarzian derivative
\begin{equation}
	\left\{ t,u\right\} = \frac{t'''(u)}{t'(u)} - \frac{3}{2} \left(\frac{t''(u)}{t'(u)}\right)^2
	\,,
\end{equation}
and used the explicit on-shell solution of the dilaton $\phi$ in \eqref{eq:LS-sol} to obtain an explicit expression for $n^{\mu}\nabla_{\mu}\phi$ arising from the bulk term \eqref{eg:bulkcontribution}.
Putting this all together we find that the full action becomes
\begin{align} \label{eq:divergingschwarzian}
	I_{\text{Lif}} = - \int du \left( \alpha E t'(u) + \frac{\epsilon^{1-\frac{1}{\alpha}}}{a t'(u)^{\frac{1}{\alpha}}} \{t,u\}_{\alpha}  \right) + \mathcal{O}(\epsilon^{3-\frac{1}{\alpha}}) \,,
\end{align}
where we defined the \emph{deformed Schwarzian} 
\begin{align}
	\{ t,u \}_\alpha :=  \left(\frac{t''(u)}{t'(u)}\right)' - \frac{1}{2\alpha} \left( \frac{t''(u)}{t'(u)} \right)^2 
	= \{t,u\} -\frac{1}{2}\left(\frac{1}{\alpha}-1\right)\left(\frac{t''(u)}{t'(u)}\right)^2 
	\, .
\end{align}
Note that for $\alpha = 1$ this coincides with the usual Schwarzian derivative. This deformed Schwarzian retains shift symmetries and rescaling symmetries, but no inversion symmetry. 

For $0 < \alpha < 1$ there is always a divergence present in equation \eqref{eq:divergingschwarzian} at order $\epsilon^{1-\frac{1}{\alpha}}$. To remedy the divergence in \eqref{eq:divergingschwarzian}, we need to rescale the black hole energy $E$ and the coordinate $t$ such that
\begin{align}
	E \to \epsilon^{1-\alpha} E\,, \hspace{1cm} t(u) \to \epsilon^{\alpha-1} t(u)
	\,,
\end{align}
which is motivated by requiring the on-shell value of \eqref{eq:divergingschwarzian} when $t(u)=u$ to be finite. 
In combination with the above we choose $u_{\text{max}}$ and $\gamma$ introduced in \eqref{eq:ranges} such that
\begin{equation}\label{eq:ranges2}
	u\sim u+\beta
	\,,
	\quad
	t(u+\beta)
	=
	t(u)
	+
	\beta
	\,.
\end{equation}	
With these rescalings we now obtain
\begin{align}
    K &= 
    a 
    + 
    \left( \frac{1}{a} \{t,u\} + E t'(u)^{1+\frac{1}{\alpha}}  \right)\epsilon^2 
    + 
	\mathcal{O}(\epsilon^{4})
    \,,  
    \\
    \sqrt{h} &=  
    \frac{1}{\epsilon}   + \mathcal{O}(\epsilon)
    \\
    e^{-2\phi} &=
   \frac{1}{t'(u)^{\frac{1}{\alpha}}} \frac{1}{\epsilon} + \mathcal{O}(\epsilon^{3})\,, 
    \\
    -2(1-\alpha)n^\mu \nabla_\mu \phi &=
    \left(\tfrac{1}{\alpha} - 1 \right) a  -(1-\alpha)
    \left(E t'(u)^{1+\frac{1}{\alpha}}  + 
    \frac{1}{2 a \alpha} \left( \tfrac{t''(u)}{t'(u)} \right)^2\right) \epsilon^2
    + \mathcal{O}(\epsilon^4)\,.
\end{align}
In this case the action is finite and taking the limit $\epsilon \to 0$ we obtain the following action
\begin{align}\label{eq:dSch1}
	I_{\text{dSch}}[t] &= -\int du \frac{1}{t'(u)^{\frac{1}{\alpha}}}\left( \alpha E t'(u)^{1+\frac{1}{\alpha}} - \frac{1-\alpha}{2 a \alpha} \left(\frac{t''(u)}{t'(u)}\right)^2 + \frac{1}{a}\left\{ t,u \right\}\right) \\
	&= -\alpha E \int t'(u) du -  \frac{1}{a} \int du \frac{1}{t'(u)^{\frac{1}{\alpha}}} \left\{ t,u \right\}_{\alpha}\,,
\end{align}
which we will refer to as the deformed Schwarzian action (dSch).
This boundary description reproduces the thermodynamics of $I_{\text{Lif}}$ when we take the saddle point approximation $t(u)=u$. We will study its loop correction in the next section.

Let us touch upon the $\alpha=1$ case and contrast it to the canonical JT result (see e.g. \cite{Maldacena:2016upp}), to emphasise the influence of choosing boundary condition \eqref{eq:boundarycond} rather than leaving it as an unspecified source. For $\alpha=1$ we establish
\begin{equation}
	I^{\alpha=1}_{\text{dSch}}[t] =
	- 
	\int du 
	~\overline{\phi}(u)
	\left[
		\frac{1}{a} \left\{ t,u\right\} 
		+
		E (t'(u))^{2}
	\right]
	\,,
	\quad
	\overline{\phi}(u)
	=
	\frac{1}{t'(u)}
	\,,
\end{equation}
where it is important to recall that we have already employed the black hole coordinate patch. By adopting the coordinate transformation to go to the Poincar\'{e} disc
\begin{equation}
	t
	=
	\sqrt{\frac{2}{aE}}\arctan \tau(u)\,,
\end{equation} 
we recover
\begin{equation}\label{eq:alphais1}
	I^{\alpha=1}_{\text{dSch}}[\tau] =
	- 
	\frac{1}{a}
	\int du 
	~\overline{\phi}(u)
		 \left\{ \tau,u\right\} 
	\,,
	\quad
	\overline{\phi}(u)
	=
	\sqrt{\frac{aE}{2}}
	\frac{1+\tau(u)^{2}}{\tau'(u)}
	\,.
\end{equation}
Let us now contrast this last results with the canonical JT outcome.
We could alternatively have started with $\alpha=1$ from the JT bulk \eqref{eq:flat-CMLS-action} with corresponding Gibbons-Hawking-York term and counter term. We could then go through the usual procedure of deriving a boundary description, with this time requiring the boundary value $e^{-2\phi}\sim \phi_{r}(u)/\epsilon$, where $\phi_{r}(u)$ will play the role of a boundary source. If we adopt a Poincar\'{e} patch with time coordinate $\tau$ we find the Schwarzian action
\begin{equation}
	I_{\text{Schwarzian}}
	\sim 
	\int du \phi_{r}(u)\{\tau,u\}
	\,,
\end{equation}
where the equations of motion for $\tau$ imply
\begin{equation}
	\phi_{r}(u)
	=
	\frac{
		\alpha
		+
		\gamma \tau(u)
		+
		\delta \tau(u)^{2}
	}{
		\tau'(u)
	}
	\,,
\end{equation}
with $\alpha$, $\gamma$, $\delta$ integration constants. To obtain the more familiar form of the Schwarzian action one can rescale $u$ such that $\phi_{r}(u)$ can be chosen a constant and effectively drops out of the action and the SL$(2,\mathbb{R})$ symmetry remains intact.
However, in order to obtain \eqref{eq:alphais1} we have chosen the integration constants $\alpha$, $\gamma$, $\delta$ such that $\phi_{r}(u)=\overline{\phi}(u)$, which is turning on a source that breaks the SL$(2,\mathbb{R})$ symmetry. 
%
%
%
%
%
%
%
\section{Quantum black holes and density of states} \label{sec:QBH}
Our goal is to use a saddle point approximation to estimate the partition function of the Lifshitz-like dilaton model. 
After the approximation is implemented, some aspects of this computation mirror computationally what happens in the determination of the JT gravity partition function as a boundary Schwarzian, but we emphasise that we \emph{approximate} the path integral, whereas for the JT gravity it has been argued in \cite{Stanford:2017thb} that the path integral of the JT Schwarzian theory is one-loop exact.
For this reason, let us posit the essence of the JT gravity approach. 
We start with specifying a regulated Euclidean $\text{AdS}_{2}$ disc and the boundary conditions for the fields at its edge, in particular for the dilaton. 
Performing the path integral over the dilaton $\phi$ fixes the Ricci scalar to be a constant and reduces the action purely to boundary terms described by the Schwarzian  action, $S_{\text{Schwarzian}}[t]$. Given the correct boundary conditions for the dilaton, treating its boundary conditions as an external source denoted by $\phi_{r}$, the Schwarzian action retains $\text{SL}(2,\mathbb{R})$ reparametrisation invariance through the boundary mode $t$, which leaves an imprint on the path integral measure over $[Dt]$. We can summarise the above words into
\begin{equation}\label{eq:partlif1}
	Z_{\text{JT}}
	=
	\int [\mathcal{D}g_{ab}][\mathcal{D}\phi]
	e^{-I_{\text{JT}}[g_{ab},\phi]}
	=
	\int_{\text{Diff}(S^{1})/\text{SL}(2,\mathbb{R})} [\mathcal{D}t]e^{-I_{\text{Schwarzian}}[t,\phi_{r}]}	
	\,.
\end{equation}
With the notation $\text{Diff}(S^{1})/\text{SL}(2,\mathbb{R})$ we mean that we integrate over full diffeomorphisms on the circle modulo the $\text{SL}(2,\mathbb{R})$ global symmetry of the Schwarzian, which reflect the isometries of $\text{AdS}_{2}$. 

For the Lifshitz-like theory we consider, integrating out the dilaton does not yield a simple result. As such we consider a saddle approximation in which we take the on-shell value for the dilaton and fix the Ricci scalar using equation of motion \eqref{eq:LS-EE}. Fixing only the Ricci scalar allows us to study a boundary mode provided we use the on-shell value of the dilaton the dilaton $\phi$ in \eqref{eq:LS-sol} to obtain an expression for $n^{\mu}\nabla_{\mu}\phi$. For us this fixes $\phi_{r}$ which, contrasting the standard JT case, does \emph{not} function as an external source but has specified dependence on $t'$.
As shown in the previous section, we obtain a deformed Schwarzian action, which has $U(1)$ global symmetry. We emphasise that within the approach we adopt, we are neglecting any possible contribution from the fact that contrasting the $JT$-setup, the dilaton cannot be integrated out exactly.

We summarise this approach as
\begin{equation} \label{eq:defSchPath}
	Z_{\text{Lif}}
	=
	\int [\mathcal{D}g_{ab}][\mathcal{D}\phi]
	e^{-I_{\text{Lif}}[g_{ab},\phi]}
	\approx
	\int_{\text{Diff}(S^{1})/\text{U}(1)} [\mathcal{D}t]e^{-I_{\text{deformed Schwarzian}}[t,\phi_{r}]}	
	\,.
\end{equation}
To evaluate the saddle approximation above we first consider fluctuations around the saddle point of the deformed Schwarzian action in subsection \ref{subsec:qpartition}. In that section we also specify the measure and evaluate the partition function. In subsection \ref{sec:dos} we study the resulting density of states and comment on these results.
\subsection{Computing the partition function of a quantum black hole} \label{subsec:qpartition}
In order to evaluate the saddle-point approximated path integral, we need to work out the action and measure and evaluate the combination. 
\paragraph{Perturbative expansion of the deformed Schwarzian action.}
The equations of motion resulting from the action
\begin{align}
	I_{\text{dSch}}[t] = -\alpha E \int t'(u) du - \frac{1}{a} \int du \frac{1}{t'(u)^{\frac{1}{\alpha}}} \left\{ t,u \right\}_{\alpha}
	\,,
\end{align}
obtained in \eqref{eq:dSch1}, are given by
\begin{align}
	\frac{d}{du} \left( \frac{\{ t,u \}_{\alpha}}{t'(u)^{1+\frac{1}{\alpha}}} \right) = 0\,. 
\end{align}
A solution is then in particular $t(u) = u$, which satisfies the ranges \eqref{eq:ranges2}, in which case we find that the classical on-shell action gives the same free energy as expected from \eqref{eq:free-energy}
\begin{equation}
	I_{\text{dSch}}[t=u] = -\alpha \beta M
	=
	\beta F
	\,.
\end{equation}

Let us now consider fluctuations around the saddle point of the deformed Schwarzian action 
\begin{align}
	t(u) = u + \epsilon(u)
	\,,
\end{align}
where $\epsilon(u + \beta) = \epsilon(u)$. At quadratic order we then find the following
\begin{align}\label{eq:quadfluct}
	I_{\text{dSch}}[t=u+\epsilon(u)] &= -\alpha \beta M  + \frac{1}{2 a \alpha} \int_0^{\beta} du \left( (1+2\alpha) \epsilon''(u)^2 + 2(1+\alpha) \epsilon'(u) \epsilon'''(u) \right) 
	+\mathcal{O}(\epsilon^{3})\nonumber
	  \\ &=  -\alpha \beta M  + \frac{1}{2 a \alpha} \int_0^{\beta} du \, \epsilon''(u)^2
	  +\mathcal{O}(\epsilon^{3})
	  \,,
\end{align}
where we performed integration by parts.
Following the approach of \cite{Saad:2019lba}, let us now expand $\epsilon(u)$ in Fourier modes
\begin{align} \label{eq:fourier}
	\epsilon(u) = \frac{\beta}{2\pi} \sum_{n \neq 0} \epsilon_n e^{- \frac{2\pi}{\beta}inu}
	\,,
\end{align}
where $\epsilon_n = \epsilon_n^{(R)} + i \epsilon_n^{(I)}$. Demanding that $\epsilon(u)$ is real therefore leads to the two following conditions $\epsilon_n^{(R)} = \epsilon_{-n}^{(R)}$ and $\epsilon_n^{(I)} = - \epsilon_{-n}^{(I)}$.  Then we find that the action can be written as
\begin{align} \label{eq:quadtratic-action}
	I_{\text{dSch}}[t=u+\epsilon(u)]  = -\alpha \beta E(\beta)  + \frac{2 \pi^2}{\beta \alpha a} \sum_{n \neq 0}  n^4 \left( \epsilon_n^{(R)} \epsilon_n^{(R)} + \epsilon_n^{(I)} \epsilon_n^{(I)}\right)\,.
\end{align}
The corresponding propagator for this action can be computed by carefully treating the $U(1)$ zero mode corresponding to $n=0$ and is given by
\begin{align}
  \left\langle \epsilon(u) \epsilon(0) \right\rangle = \frac{\beta \alpha a}{2 \pi} \left( -\frac{1}{24} \left( u-\pi \right)^4 + \frac{\pi^2}{12}(u-\pi)^2 - \frac{7 \pi^4}{360} \right)\,.
\end{align}

\paragraph{Obtaining the measure.}
In order to evaluate the one-loop correction to the saddle-point approximation in \eqref{eq:defSchPath} we first need to specify the measure $\left[ \mathcal{D}t \right]_{\text{Lif}}$ which the path integral will be computed with. Let us first recall the outlines of the calculation of the measure for JT before specifying the form for our Lifshitz-like dilaton model.

Due to the fact that Jackiw-Teitelboim gravity is $\text{SL}(2,\mathbb{R})$ invariant the measure $[\mathcal{D}t]_{\text{JT}}$ explicitly does not involve any modes which give rise to Killing vectors corresponding to the $\text{SL}(2,\mathbb{R})$ isometries of $\text{AdS}_2$. The measure for JT has been computed by \cite{Stanford:2017thb, Moitra:2021uiv} and is
\begin{align}
	[\mathcal{D}t]_{\text{JT}} = \prod_u \frac{dt(u)}{t'(u)} =  \prod_{n\geq 2} 4\pi n (n-1)(n+1) d \epsilon_{n}^{(R)} d \epsilon_n^{(I)}\,.
\end{align}
where we used the saddle point approximation $t(u) = u + \epsilon(u)$ with $\epsilon(u)$  given by equation \eqref{eq:fourier}. Let us emphasise again that the product is taken for $n \geq 2$ since $n=-1, n= 0$ and $n = +1$ specifically correspond to removed zero modes due to the $\text{SL}(2,\mathbb{R})$ isometries transformations of the bulk solution.
Due to the fact that in the Lifshitz-like model we consider, the dynamical dilaton breaks the $\text{SL}(2,\mathbb{R})$ symmetry down into $U(1)$ where the only Killing vectors corresponding to the $U(1)$ isometries of $\text{AdS}_2$ we are lead to consider the following measure
\begin{align} \label{eq:lifshitz-measure}
	[\mathcal{D}t]_{\text{Lif}} = \prod_{n \geq 1} 4\pi n d\epsilon_{n}^{(R)} d \epsilon_n^{(I)}\,.
\end{align}
A few remarks are in order. Let us first note that the most drastic change in the evaluation of the partition function now comes from the product also including the contribution from the $n = \pm 1$ modes, since the classical solution of the deformed Schwarzian only has one zero mode that can be removed, corresponding to $U(1)$ isometry. Let us also remark that the exact power of $n$ does not matter so much in the evaluation of the path integral since it only changes the numerical constant in front of the partition function and not its thermodynamical behaviour. Here we have chosen the lowest degree polynomial. 
\paragraph{Evaluating fluctuations around saddle points.}
Having stipulated the form of the measure for the Lifshitz-like dilaton gravity model in equation \eqref{eq:lifshitz-measure} along with the quadratic action in equation \eqref{eq:quadtratic-action} we may now compute the one-loop saddle-point corrected partition function of the Lifshitz-like dilaton models as follows
\begin{align}
	Z_{Q}(\beta) &\approx \int_{\text{Diff}(S^1)/U(1)} \left[ \mathcal{D} t \right]_{\text{Lif}} \exp(\alpha E(\beta) \int t'(u)du + \frac{1}{a} \int du \frac{\left\{ t,u \right\}_{\alpha}}{t'(u)^{\frac{1}{\alpha}}}) \\ &=e^{\alpha \beta E(\beta)} \prod_{n \geq 1} 4\pi n \int d\epsilon_n^{(R)} d\epsilon_{n}^{(I)} e^{- \frac{2\pi^2n^4}{\beta \alpha a}\left( \epsilon_n^{(R)}  \epsilon_n^{(R)}  +  \epsilon_n^{(I)}  \epsilon_n^{(I)}   \right)} \\ &= e^{\alpha \beta E(\beta)} \prod_{n \geq 1}  \frac{2 \alpha \beta a}{n^3} = \frac{1}{4 \pi^{3/2} \sqrt{a \alpha \beta}} e^{\alpha \beta E(\beta)}\,. \label{eq:partitionfunction}
\end{align}
where in the last step we used zeta function regularisation, namely that
\begin{align}
	\prod_{n\geq 1} \frac{2 \alpha \beta a}{n^3} 	
= \exp[ \lim_{\epsilon \to 0} \partial_\epsilon \left( \left( 2 \alpha \beta a \right)^{\epsilon} \zeta(3\epsilon)\right)  ] = 
	\frac{1}{4 \pi^{3/2} \sqrt{a \alpha \beta}}
	\,,
\end{align}
where $\zeta(3\epsilon):=\sum_{n \geq 1}n^{-3 \epsilon}$.

The quantum corrected free energy $F_{Q}$ reads
\begin{equation}\label{eq:Cdef}
	-\beta F_{Q}
	:=
	\log Z(\beta)
	=
	\frac{ 
	\alpha C
	 }{\beta^{\frac{1}{\alpha}}}
	+
	\frac{1}{2}
	\log \frac{1}{a \beta}
	+
	\text{...}
	\,,
	\quad
	C
	:=
	\left(\frac{2 \pi}{1+\alpha}\right)^{\frac{1+\alpha}{\alpha}}
	\left(\frac{2\alpha}{a}\right)^{\frac{1}{\alpha}}
	\,,
\end{equation}
where the dots denote temperature independent constants. The logarithm is due to taking into account the loop corrections. We note that the factor $1/2$ reflects the single generator of the symmetry group. In the $\text{SL}(2,\mathbb{R})$ case, for example, this factor is $3/2$, reflecting its three generators. Furthermore, as we lower the temperature, from $\beta\sim 1/a$ onwards we expect quantum corrections to take over. In fact, solving $F_{Q}(\beta_{*})=0$ yields the temperature $\beta_{*}>0$ at which one expect non-perturbative effects to the saddle point approximation to arise.

\subsection{Density of states}\label{sec:dos}
The density of states classifies the energy spectrum of a system, such as the quantum black hole we are studying. We can obtain the density of states through the partition function. 
The partition function $Z$ is related to the density of states $\rho$ in the following manner
\begin{align}
	Z(\beta) = \int_0^\infty dE \, \rho(E) e^{-\beta E}\,,
\end{align}
in other words, $Z(\beta)$ is the Laplace transform of the density of states, $\rho(E)$. Traditionally we may compute the density of states from the partition function $Z(\beta)$ by an inverse Laplace transform 
\begin{align}
	\rho(E) = \frac{1}{2\pi i} \int_{c - i\infty}^{c + i \infty} Z(\beta) e^{\beta E} d\beta \, ,
\end{align}
where the integral is performed along a vertical line in the complex plane and the constant $c \in \mathbb{R}$ is a sufficiently large so that we may enclose all poles of the partition function with a semi-circle in the left half-plane. In practice this can often be evaluated with the Cauchy residue theorem.
\paragraph{Classical spectrum.}
To contrast the quantum spectrum, let us first study the classical black hole spectrum by considering just the saddle point approximation:
\begin{align} \label{eq:clDOS}
	Z_{\text{cl}}(\beta) \approx e^{-\mathcal{I}_{\text{classical}}} =  e^{\alpha \beta E(\beta)} = e^{\alpha C \beta^{-\frac{1}{\alpha}}}\,.
\end{align}
Here we have made use of the thermodynamical relationship from equation \eqref{eq:free-energy} and the defintion of $C$ in \eqref{eq:Cdef}. By expanding the exponential function in equation \eqref{eq:clDOS} in a formal series and taking the inverse Laplace transform of each term it follows that the classical density of states is given by
\begin{align} \label{eq:densityofStates}
	\rho_{\text{cl}}(E) = \delta(E) + \frac{1}{E} \sum_{n=1}^{\infty} \frac{\left( \alpha C E^{\frac{1}{\alpha}} \right)^n}{n! \, \Gamma[\frac{n}{\alpha}]} \, ,
\end{align}
where notably the $\delta$ function arises as the inverse Laplace transform of the first term in the formal series. For general values of $\alpha$ we are unable to simplify the formal series further. However, for $\alpha=1$ we find
\begin{equation}
	\rho^{\alpha=1}_{\text{cl}}(E)
	=
	\delta(E)
	+
	\sqrt{\frac{\alpha C}{E}}
	I_{1}(2\sqrt{\alpha C E})
	\,,
\end{equation}
where $I_{1}$ denotes a modified Bessel function of the first kind, see e.g. \cite{Hernandez-Cuenca:2024icn} for a coinciding result in the context of super symmetric Schwarzian theories.

To get some insights into the expression for $\rho_{\text{cl}}(E)$ for general $\alpha$, we will investigate the large and small energy regimes.
In the large energy approximation $C E^{\frac{1}{\alpha}} \gg 1$ we have
\begin{align} \label{eq:clea}
	\rho_{\text{cl}}(E) \approx \frac{C^{\frac{\alpha}{2(1+\alpha)}}}{E^{\frac{1+2\alpha}{2+2\alpha}}\sqrt{2 \pi (1+\frac{1}{\alpha})}}  e^{(1+\alpha) C^{\frac{\alpha}{1+\alpha}} E^{\frac{1}{1+\alpha}}}
	\sim
	e^{S(E)}
	\,,
\end{align}
where $S(E)$ denotes the Bekenstein-Hawking entropy, as is expected for large enough black holes. Let us now explore the small energy regime of \eqref{eq:densityofStates}. In the small energy approximation $C E^{\frac{1}{\alpha}} \ll 1$ we find that
\begin{align}\label{eq:lowenergyclassical}
	\rho_{\text{cl}}(E) \approx 
	\delta(E)
	+
	\frac{\alpha C E^{\frac{1}{\alpha}-1}}{\Gamma[\frac{1}{\alpha}]}\,.
\end{align}
Despite the fact that the density of states appearing above blows up in the low energy limit then this does not mean that it is pathological. What matters more is whether or not the total concentration of available states in the system between any two energies $E_1$ and $E_2$, where $E_{1}<E_{2}$, is finite, i.e.~whether
\begin{align}\label{eq:energybands}
	N_{\text{cl}}^{\text{tot}}(E_1, E_2) = \int_{E_1}^{E_2} \rho_{\text{cl}}(E) \, dE < \infty
	\,,
\end{align}
which is the case for all values of $0<\alpha\leq1$. In the low energy approximation of the density of states of the classical saddle \eqref{eq:lowenergyclassical}, we furthermore point out that the power of $E$ is always positive. 
As such, away from $\alpha=1$, the behaviour of the density of states is such that there is a fall-off as $E\to0$.
Finally, it is significant that for $0<\alpha\leq1$
\begin{equation}
	N_{\text{cl}}^{\text{tot}}(E_{1}=0,E_{2}\to0)\to 1
	\,,
\end{equation}
because reinstating the extremal entropy $S_{0}$ of the higher dimensional black hole reveals that the ground state of the classical saddle has a large degeneracy of $\exp S_{0}$. This is expected to disappear when corrections are added, for non-super symmetric theories.
\paragraph{Quantum corrected spectrum.}
Let us now contrast density of states of the classical saddle with the one-loop corrected partition function.
By taking the inverse Laplace transform of the one-loop corrected partition function given by equation \eqref{eq:partitionfunction} we find
\begin{align} \label{eq:qdensityofStates}
	\rho_{\text{Q}}(E) = \frac{1}{4 \pi^{3/2} \sqrt{a \alpha E}} \sum_{n=0}^{\infty} \frac{(\alpha C E^{\frac{1}{\alpha}})^n}{n! \, \Gamma[\frac{1}{2} + \frac{n}{\alpha}]} \, .
\end{align}
For the case of $\alpha=1$ we find the closed expression
\begin{equation}
	\rho_{\text{Q}}^{\alpha=1}(E)
	=
	\frac{
		\cosh 2\sqrt{C E}
	}{
		4\pi^{2}\sqrt{a E}
	} 
	\,,
\end{equation}
which we will discuss further in the next subsubsection and contrast this result with the more usual JT result where the $\text{SL}(2,\mathbb{R})$ remains intact.
For general values of $\alpha$ we can gain more insights into the density of states of the quantum black hole through approximations. We consider the large energy approximation $C E^{\frac{1}{\alpha}} \gg 1$, for which we find that the quantum effects are washed out and the classical saddle point answer is recovered:
\begin{align} \label{eq:qlea}
	\rho_{\text{Q}}(E) \approx \frac{1}{4 \sqrt{2} \, \pi^2 \sqrt{(1+\alpha) a E}}  e^{(1+\alpha) C^{\frac{\alpha}{1+\alpha}} E^{\frac{1}{1+\alpha}}}
	\sim
	\rho_{\text{cl}}(E)
	\,,
\end{align}
which reinforces that the quantum corrections do not influence the high energy regime.

\begin{figure}[H]
\centering
\begin{subfigure}{.45\textwidth}
  \centering
  \includegraphics[width=.8\linewidth]{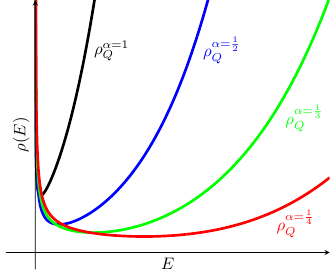}
  \caption{The quantum corrected density of states \\ $\rho_Q(E)$ for different values of $\alpha$ according to \\equation \eqref{eq:qdensityofStates}. Note the small energy \\ dependence $\rho_Q \sim \frac{1}{\sqrt{E}}$ independent of $\alpha$.}
  \label{fig:qdsub1}
\end{subfigure}
\begin{subfigure}{.45\textwidth}
  \centering
  \includegraphics[width=.8\linewidth]{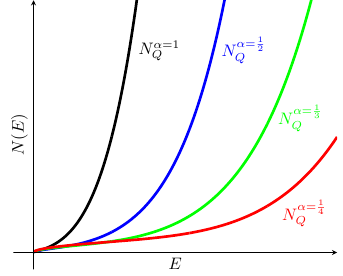}
  \caption{The number of states, $N(E)$ below a certain energy $E$ for different values of the parameter $\alpha$ according to equation \eqref{eq:energybands}. Note the lack of an exponentially large ground state.}
  \label{fig:qdsub2}
\end{subfigure}
\caption{Highlighting the low energy behaviour of the quantum corrected density of states.}
\label{fig:qNdos}
\end{figure}

\begin{figure}[H]
\centering
\begin{subfigure}{.45\textwidth}
  \centering
  \includegraphics[width=.8\linewidth]{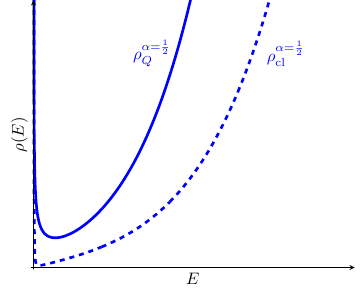}
  \caption{Comparing the classical density of states \\ \eqref{eq:clDOS} to the quantum corrected \eqref{eq:qdensityofStates} for \\ the specific value  $\alpha = 1/2$.}
  \label{fig:comp1}
\end{subfigure}
\begin{subfigure}{.45\textwidth}
  \centering
  \includegraphics[width=.8\linewidth]{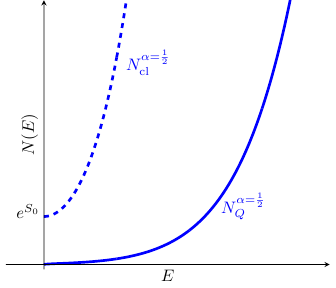}
  \caption{Comparing classical and quantum corrected expressions for  $\alpha = 1/2$. Note the absence of a degenerate ground state for the quantum correction.}
  \label{fig:comp2}
\end{subfigure}
\caption{Comparing the classical and quantum corrected density of states.}
\label{fig:comp12}
\end{figure}

To get a better idea of the quantum mechanical effects we now consider the small energy regime. 
In the small energy limit for $CE^{\frac{1}{\alpha}} \ll 1$ this means that
\begin{align}\label{eq:densityquantum}
	\rho_{Q}(E) \approx \frac{1}{4 \pi^{2} \sqrt{a \alpha E}}
	\,.
\end{align}
Contrasting the classical saddle point \eqref{eq:lowenergyclassical}, the density of states in the quantum case rapidly grows for small energies instead of tapering off towards zero. However, when integrating this density we find that finite behaviour for small energies. In, fact we find
\begin{align}\label{eq:energybands}
	N_{\text{Q}}^{\text{tot}}(E_1, E_2) = \int_{E_1}^{E_2} \rho_{\text{Q}}(E) \, dE
	\,,
	\quad
	N_{\text{Q}}^{\text{tot}}(E_1=0, E_2\to0)\to0
	\,, 
\end{align}
for all $0<\alpha\leq1$. This means that even if we were te reinstate the extremal degeneracy $\exp S_{0}$, the ground state will have no degeneracy when quantum corrected (see \autoref{fig:qNdos} and \autoref{fig:comp12}).

The low energy behaviour of the dispersion relation can be reproduced from simple free theories. Namely, let us consider a $D$-dimensional theory with a free Lifshitz-type dispersion relation $E\sim |k|^{z}$, where $k$ is the momentum. 
The density of states then becomes
\begin{equation}
	\rho(E)
	:=
	\frac{
		d\Omega(k(E))
	}{
		d|k|
	}
	\sim
	\frac{1}{\sqrt{E}}
	\,,
	\quad
	\Leftrightarrow
	\quad
	z
	=
	2(D-1)
	\,,
\end{equation}
where we used that the total number of states contained in a sphere in momentum space scales as $\Omega(k)\sim |k|^{D-1}$. We find that this value of $z$ satisfies $D\leq z+1$, which for models of Lifshitz models is reported to behave differently from $D<z+1$, see e.g. \cite{Sybesma:2015oha,Keranen:2016ija}. We point out that in particular the behaviour of the density of states matches that of a $1+1$-dimensional theory with a non-relativistic dispersion relation.

\paragraph{Comparison with the usual JT gravity approach}
It is important to stress that $\alpha=1$ corresponds to a boundary description of JT gravity with a source turned on that breaks the $\text{SL}(2,\mathbb{R})$ down to U(1) for the classical saddle. This contrasts the usual approach in which such a source is not considered and the classical saddle retains its $\text{SL}(2,\mathbb{R})$ invariance. In the usual approach the density of states of the quantum corrected saddle becomes proportional to a hyperbolic sine of energy \cite{Saad:2019lba}, which goes to zero as energy vanishes. The resulting conclusion is that there is a very low degeneracy for these black holes at low energy. In our case, for any $\alpha$, the opposite is true, as for low energy the density of state grows with the inverse square root of energy, see \eqref{eq:densityquantum}.

Furthermore, let us summarize succinctly that for $\alpha = 1$ we found that the partition function and density of states were respectively given by
\begin{align}
	Z_Q^{\alpha = 1}(\beta) = \frac{1}{4 \pi^{3/2} \sqrt{a \beta}} e^{C \beta^{-1}}, \hspace{1cm} 	\rho_{\text{Q}}^{\alpha=1}(E)
	=
	\frac{
		\cosh 2\sqrt{C E}
	}{
		4\pi^{2}\sqrt{a E}
	} 
	\,.
\end{align}
This result coincides with the trumpet result for JT in equation (126) of Saad-Shenker-Stanford \cite{Saad:2019lba} where notably the $\text{SL}(2,\mathbb{R})$ symmetry is also broken down to $U(1)$. This same form has also appeared in the context of  $\mathcal{N}=1$ supersymmetric JT gravity.\footnote{see e.g.~equation (3.46) of \cite{Stanford:2017thb}, (6.4) of \cite{Mertens:2017mtv}, (5.4)-(5.8) of \cite{Stanford:2019vob}, or (2.25) of \cite{Hernandez-Cuenca:2024icn}.}

\section{Discussion and Outlook}\label{sec:discussions}

In this paper we have computed a one-loop correction to the partition function of 4d near-extremal charged Lifshitz black holes that arise in low energy effective actions of string theory. This has been accomplished by dimensionally reducing the four-dimensional black hole to an effective two-dimensional dilaton theory of gravity which is asymptotically $\text{AdS}_2$ and arises in the near-extremal charge limit of the Lifshitz black hole. By imposing wiggling boundary conditions we derived a boundary action formulation that resembles the Schwarzian action, with the key exception that our boundary action does not retain the $\text{SL}(2,\mathbb{R})$, but breaks it down to $U(1)$. 
A key difference with the JT case is that we perform a saddle point approximation of the dilaton to obtain the boundary description.
We employed the boundary action  in order to compute the quantum corrected partition function of these Lifshitz black holes. Our analysis shows that for small temperatures the quantum one-loop corrected partition function is given by
\begin{align}
	Z_{\text{BH}}(T) \sim \sqrt{T} e^{S_0} + \emph{higher order terms} \,,
\end{align}
where $S_{0}$ is the extremal entropy.
This indicates that including quantum corrections to the black hole entropy removes the exponential ground-state degeneracy observed in the semiclassical analysis. This corresponds with the results reported for Reissner-Nordström and Kerr-Newman black holes in recent work \cite{Iliesiu:2022onk, Banerjee:2023quv, Kapec:2023ruw, Rakic:2023vhv}, except that we find that the prefactor is $\sqrt{T}$ instead of $T^{3/2}$. This difference in prefactor stems from the lack of symmetries present at the horizon of these Lifshitz black holes, which merely retain $U(1)$ symmetries instead of the full $\text{SL}(2,\mathbb{R})$ symmetries. There is an intricate connection between the amount of symmetries present in the near-extremal region of the black hole and the prefactor of the one-loop corrected partition function. 

The one-loop corrected partition function $Z_{\text{BH}}(T)$, which we computed vanishes in the low temperature limit, indicating that the corresponding density of states approaches zero as energy decreases, thereby removing the previously observed exponential ground-state degeneracy. The expression that we obtained for the quantum one-loop corrected density of states in equation \eqref{eq:qdensityofStates} behaves as $\rho_Q(E) \sim 1/\sqrt{E}$ for small energies indicating that the quantum corrected density of states blows up for small energies, i.e.~$\rho(E) \rightarrow \infty$ as $E \to 0$.  This same low energy behaviour of the density of states is encountered in several non-relativistic theories, the most elementary being the one-dimensional infinite potential well. Note that integrating over a small band of energy smoothes out this blowing up behaviour and the total number of states does go to zero as $E\to0$, as highlighted pictorially in \autoref{fig:comp12}. In contrast the classical spectrum clearly exhibits an exponentially large ground state degeneracy. 
 
In our approach we of course do not capture the full four-dimensional near-extremal one-loop corrections: we have ignored vector and gauge modes, which our two-dimensional approach fails to capture (we only capture the tensor mode). 
It would be interesting to see if we could capture those other contributions using the recently developed methods for computing the logarithmic corrections to the black hole entropy in \cite{Iliesiu:2022onk, Banerjee:2023quv, Kapec:2023ruw, Rakic:2023vhv, Banerjee:2010qc, Banerjee:2011jp, Sen:2012kpz, Sen:2012cj}, which rely on the fact that the normalised basis of zero mode deformations of $\text{AdS}_2$ is well known and is utilised to simplify the computation of the logarithmic correction. We were unable to perform the same feat for $\text{Lif}_{2}$ due to the technical difficulty of obtaining the zero modes in that spacetime, which we leave to future work. Another potential contribution to the path integral that we neglected are possible bulk fluctuation contributions.

As a conclusion we note that it would be interesting to pursue whether there exists a matrix model dual these Lifshitz-like dilaton theories, similar to JT gravity. This seems to be a non-trivial task, as the action does not appear to be a deformation of JT gravity \cite{Witten:2020wvy}. Finally, in the case of the boundary description of JT gravity it has been argued that the partition function of double-scaled SYK for low energies agrees with the Schwarzian result \cite{Saad:2019lba,Mertens:2017mtv,Lam:2018pvp,Berkooz:2018jqr}. It would be interesting to construct a quantum mechanical dual to our constructed boundary description that breaks down $\text{SL}(2,\mathbb{R})$ to $U(1)$.
\section*{Acknowledgements}
We thank Alex Frenkel, Friðrik Freyr Gautason, and Diego Hidalgo for discussions and insightful comments. 
WS thanks Matt Blacker, Alejandra Castro, and Chiara Toldo for collaboration on related topics.
MBH and WS acknowledge the Icelandic Research Fund Grant 228952-053.
WS also acknowledges support of the Simons Foundation through the INI-Simons Postdoctoral Fellowship.

\newpage
\appendix

\bibliographystyle{JHEP}
\bibliography{LS}

\end{document}

%% file: puzzle-diagram.tex
\scalebox{0.7}{
\begin{tikzpicture}
  \draw[thick,-Stealth] (-0.1,0) -- (8,0) node[below] {$T$};
  \draw[thick,-Stealth] (0,-0.1) -- (0,8) node[right] {$E(T)$};

  \draw[thick,dotted] (4,4) -- (4,0) node[below] {$T_{\text{gap}}$};
  \draw[thick,dotted] (4,4) -- (0,4) node[left] {$E_{\text{gap}}$};

  \draw[scale=1, domain=0:6, thick, smooth, variable=\x, blue, samples=1] plot ({\x}, {\x});
  \draw[scale=1, thick, domain=0:5, smooth, variable=\x, violet, samples=31] plot ({\x}, {0.9*\x + 0.25*\x^(3/2)});
  \draw[scale=1, domain=0:6, smooth, thick, variable=\x, red, samples=31] plot ({\x}, {0.5*\x^(3/2)});

  \node[blue] at (6.5,5) {$E_{\text{Hawking}} \sim T$};
  \node[red] at (7,6.8) {$E_{\text{BH}} \sim T^{2}$};
  \node[violet] at (3,6.8) {$E_{\text{BH}} \sim \frac{3}{2}T + T^{2}$};

  \node[circle,fill,inner sep=1.3pt] at (4,4) {};

  \node at (1.5,-0.25) {\scriptsize \emph{semiclassical}};
  \node at (1.5,-0.65) {\scriptsize \emph{description breaks down}};
  \node at (-1.5,0.25) {\scriptsize Extremal Black Hole};
  \node at (2.9,8) {\scriptsize (Energy above extremality)};
  \node at (-1,-0.25) {$T = 0$};
  \node[violet] at (2.8,6.25) {\scriptsize quantum correction};
\end{tikzpicture}
  }